\documentclass[aps, twocolumn]{revtex4}

\usepackage[T1]{fontenc}
\usepackage{graphicx}
\usepackage{amsmath, amsthm, amssymb, amsfonts}
\usepackage{mathtools}
\usepackage{verbatim}
\usepackage{dcolumn}
\usepackage{bm}
\usepackage{epsf}
\usepackage{color}
\usepackage[colorlinks=true,citecolor=blue,linkcolor=red,urlcolor=myBlue]{hyperref}
\usepackage{xcolor}
\usepackage{physics}
\usepackage{dsfont}
\usepackage{tikz}
\usepackage{multirow}
\usepackage{cleveref}
\usepackage{siunitx}
\usepackage{xurl}
\usepackage[autostyle]{csquotes}
\usepackage{algorithm, algpseudocode}
\usepackage{subcaption}
\usepackage{booktabs}
\usepackage[dvipsnames]{xcolor}
\definecolor{myBlue}{RGB}{31,119,180}
\definecolor{myOrange}{RGB}{255,127,14}
\definecolor{myGreen}{RGB}{44,160,44}
\definecolor{myRed}{RGB}{214,0,0}
\definecolor{myPurple}{RGB}{148,103,189}
\definecolor{myDarkBlue}{RGB}{0, 0, 225}

\begin{document}

\title{Improving the efficiency of QAOA using efficient parameter transfer initialization and targeted-single-layer regularized optimization with minimal performance degradation}

\author{Shubham Patel}
\email{shubhampatel.research@gmail.com}
\affiliation{Department of Physics \& Astrophysics, University of Delhi, Delhi-110007.}

\author{Utkarsh Mishra}
\email{umishra@physics.du.ac.in}
\affiliation{Department of Physics \& Astrophysics, University of Delhi, Delhi-110007.}

\begin{abstract}
Quantum approximate optimization algorithm (QAOA) have promising applications in combinatorial optimization problems (COPs). 
We investigated  the MaxCut problem in three different families of graphs using QAOA ansats with parameter transfer initialization
followed by targeted-single-layer optimization. For $3$-regular (3R), Erdős–Rényi (ER), and Barabási–Albert (BA) graphs, the parameter transfer approach achieved mean approximation ratios of $0.9443$ for targeted-single layer optimization ($r_s$) as compared to $0.9551$ of full optimization ($r_f$). It represents $98.88$\% optimal performance, with $8.06$$\times$computational speedup in unweighted graphs. But, in weighted graph families, optimal performance is relatively low ($<90\%$) for higher nodes graph, suggesting parameter transfer followed by targeted-single-layer optimization is not ideal for weighted graph families, however, we find that for some weighted families (weighted 3-regular) this approach works perfectly. In $8.92$\% test cases, targeted-single-layer optimization outperformed the full optimization ($r_s\geq r_f$), indicating that complex parameter landscape can trap full optimization in sub-optimal local minima. To mitigate this inconsistency, ridge ($L_2$) regularization is used to smoothen the solution landscape, which helps the optimizer to find better optimum parameters during full optimization and reduces these inconsistent test cases from $8.92\% \to 3.81\%$. This work demonstrates that efficient parameter initialization and targeted-single-layer optimization can improve the efficiency of QAOA with minimal performance degradation.
\end{abstract}

\maketitle

\section{Introduction} \label{sec:introduction}
Following Feynman, several theoretical and experimental efforts have been made to harness fundamental principles of quantum mechanic in a controlled way to perform  computation that are widely considered to be intractable in classical computers~\cite{feynman2018simulating, Deutsch, Deutsch-Jozsa, Bernstein, 365700, grover1996fastquantummechanicalalgorithm}. Combinatorial optimization problems (COPs)~\cite{SCHRIJVER20051}, such as the Maximum-Cut (MaxCut)~\cite{Festa01012002, Commander2009} problem, are NP-hard~\cite{NP_completeness}, implying that they cannot be solved in polynomial time on classical computers for large-scale instances. The goal of the MaxCut problem is to partition the vertices of a graph into two disjoint subsets such that the total weight of the edges crossing the partition is maximized. The well-known quantum algorithm to find an approximate solution to the MaxCut problem is the \enquote{Quantum Approximate Optimization Algorithm} (QAOA)~\cite{farhi2014quantumapproximateoptimizationalgorithm}.

QAOA operates by applying $p$ layers of cost ($U_C (\gamma_i)$) and mixer ($U_M(\beta_i)$) unitary alternatively, with each new layer improving the solution quality. 
A quantum-classical feedback loop working as a bridge between quantum ansatz and classical optimizer is responsible for providing updated parameters from the optimizer to ansatz at each iteration.
Optimizable parameters increase linearly with the circuit depth ($2p$), however such optimization may become more complicated due to exponentially increasing ($2^n$) solution landscape with input problem size ($n$). The solution landscape in MaxCut is generally  non-convex which can trap the optimization process in sub-optimal local minima, especially for the weighted graph families~\cite{PhysRevA.109.062602, meng2025conditionaldiffusionbasedparametergeneration}.

The quality of MaxCut solution is measured by the approximation ratio ($r$), defined as ratio of cost function obtained by QAOA, $\braket{H_C}$ (i.e., the expectation value of the quantum cost Hamiltonian), and the exact MaxCut value ($C_{max}$).  It is established that as $p\to \infty$, QAOA tends to yield exact value of $r=1$~\cite{farhi2014quantumapproximateoptimizationalgorithm}. For $p=1$, the approximation ratio, $r  \geq 0.69$ using QAOA ansatz. To obtain quantum advantage in QAOA, one need to increase circuit depth $p$ such that $r\geq 0.878$, which was achieved by the best known classical algorithm for MaxCut,~\enquote*{Goemans-Williamson MaxCut approximation algorithm (GW)}~\cite{GW} capable of achieving approximate MaxCut solution in polynomial time ($r_{GW}\geq 0.878$).


\begin{figure*}
	\centering
	\includegraphics[width=1\linewidth]{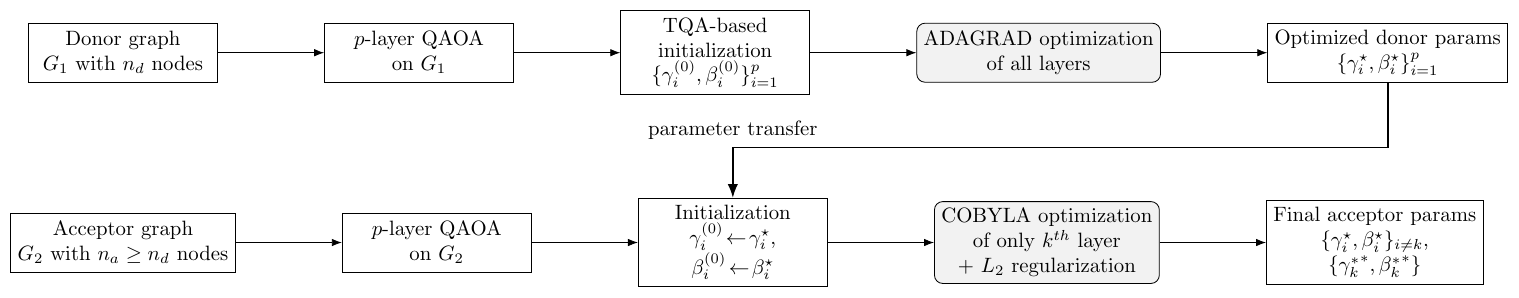}
	\caption{Schematic diagram demonstrating working of QAOA with parameters transfer initialization from a simple instance $G_1$($n_d$ nodes) to more complex instance $G_2$($n_a\geq n_d$) followed by targeted-single layer ($k^{th}$ layer) optimization with $L_2$ regularization.}
	\label{PTSLO_Schematic}
\end{figure*}


In deep learning, parameters transfer is a very popular tool \cite{tan2018surveydeeptransferlearning,zoph2020rethinkingpretrainingselftraining,yosinski2014transferablefeaturesdeepneural,rebuffi2017learningmultiplevisualdomains, 9605328, 10.1145/3584706, sakai2024linearly}, where parameters between similar instances are transferable to each other. 
This technique is implemented in QAOA by using optimized parameters of simple instances as an initialization for similar complex instances.
The parameters transfer followed by targeted-single-layer optimization can further improve the approximation ratio \cite{cjjm-87gl}, as the transfer parameters provides a near optimal choice for the initialization.


We obtained approximate MaxCut values using parameter transfer and targeted-single layer optimization, combined with regularization  techniques, across three graph families: 3-Regular, Erdős–Rényi, and Barabási–Albert graphs. For parameters transfer, $n_d = 8$-nodes graphs (donor) are used from each family whose parameters are initialized using TQA~\cite{Sack_2021} ($t=0.75$) initialization and optimized through Adagrad~\cite{JMLR:v12:duchi11a} ($lr = 0.1$, $\epsilon = 10^{-8}$, $iterations=100$) optimizer for full optimization. We consider $50$ graphs for each nodes ($8, 10, \ldots, 24$) and  identified the single layer ($p_k$) producing best $r_s$, termed as the targeted-single-layer. 
In acceptor graph, each instance is taken 20 times to obtain average value, $r$. Our result demonstrates that with this approach, QAOA requires very few optimization-steps ($N$) and reduced optimization-time ($\tau$) compared to full optimization. Moreover, it always outperforms the full optimization when compared with efficiency in unweighted cases and sometimes in weighted cases also.  
This technique preserves the advantage of a deep QAOA ansatz and reduces the optimization process by optimizing only a single layer.

The paper is organized as follows. Section~\ref{sec:methods} outlines the methods used, including parameter transfer, targeted-single layer optimization, and regularization. Section~\ref{sec:results} presents the results obtained for the three graph families. Finally, Section~\ref{sec:conclusion} concludes the paper with a summary of findings and future directions.

\section{Methods} \label{sec:methods}
In this section, we outline 
parameter transfer technique integrated in QAOA for initialization followed by targeted-single-layer optimization to further improve the solution. The schematic diagram for working of this algorithm is shown in Fig.~\ref{PTSLO_Schematic}. To initialize donor graph, we used Trotterized Quantum Annealing (TQA), discussed in the following section. We also discuss various regularization methods to make the complex cost function landscape smooth.

\subsection{Trotterized Quantum Annealing (TQA)}
TQA protocol provides the initialization of QAOA parameters in terms of discrete time steps. It is shown that this initialization allows the optimizer to avoid the issue of suboptimal solution~\cite{Sack_2021}. For a depth-$p$ QAOA ansatz, the parameters initialization by TQA is given by:
\begin{equation}
	\gamma_i = \frac{i}{p}\Delta t, \ \ \ \beta_i = \left(1-\frac{i}{p}\right)\Delta t.
\end{equation}
Here, $i\in \{0, 1, .., p-1\}$, making the $\gamma_i$ as linearly increasing and $\beta_i$ as linearly decreasing parameters over the increasing circuit depth, $p$. This is akin to the quantum annealing process where parameters are tuned in a controlled way to find the ground state energy of complex Hamiltonian~\cite{crosson2021prospects}.
 We described parameterized quantum ansatz for QAOA in Appendix~\ref{app_qaoa}.

\subsection{Parameters Transfer Learning}
\begin{figure*}
		\includegraphics[width=\textwidth]{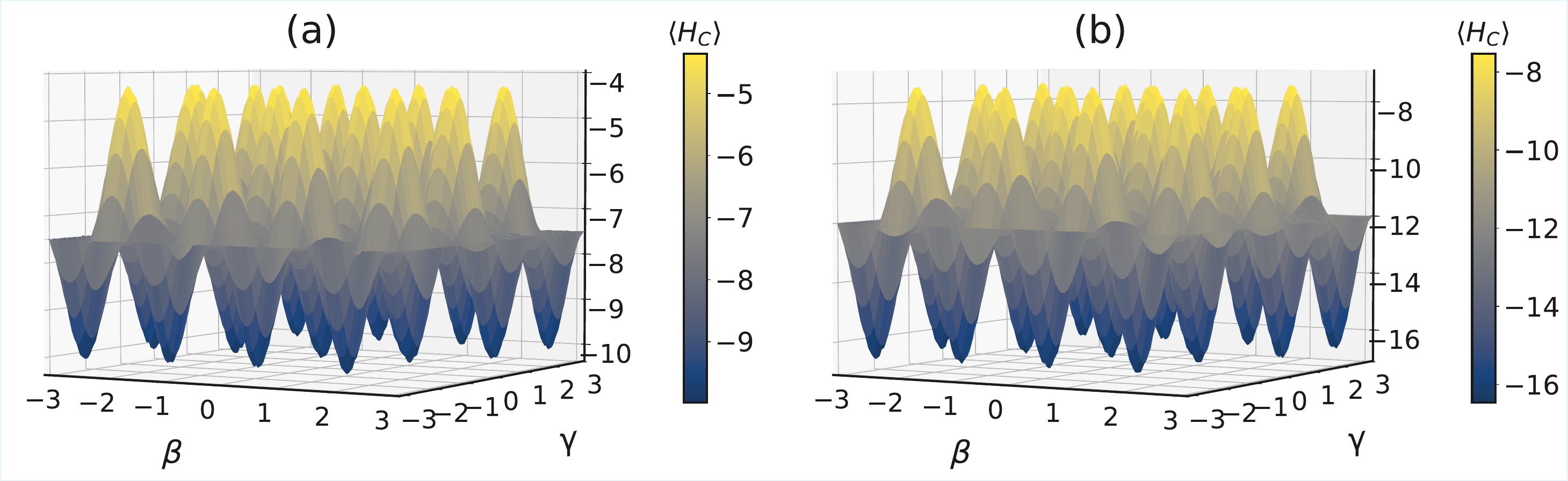}
		\caption{\textbf{Parameters solution space:} (a)for $10n-u3R$ and (b) for $16n-u3R$ graph using $p=1$ QAOA ansatz.}
		\label{fig:Landscape_8-12}
	\end{figure*}
\begin{figure*}
    
		\includegraphics[width=\textwidth]{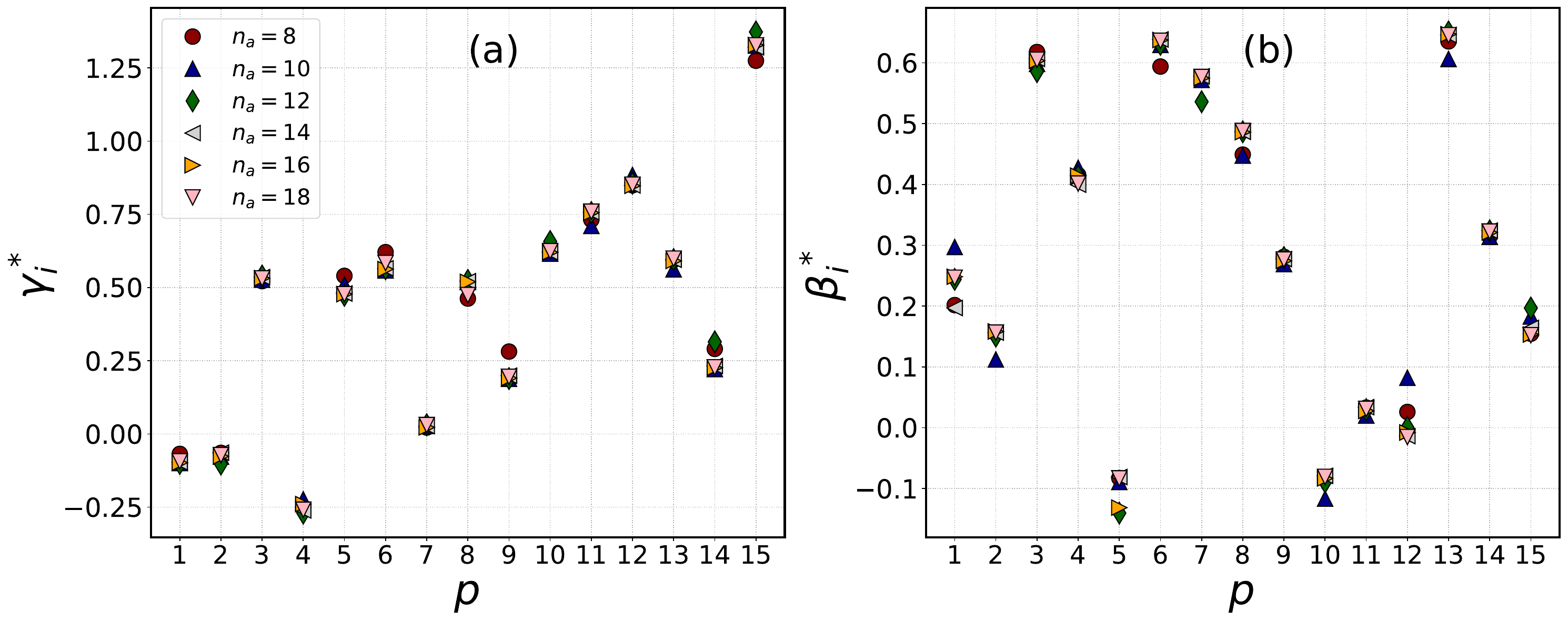}
		\caption{\textbf{Clustering of parameters:} (a) Plot of optimal $\gamma^{*}_i$ as a function of increasing layers $p$ for similar instances of u3R family.(b) Similar plot for $\beta^{*}_{i}$.}
        \label{fig:optimized_par}
\end{figure*}
Recent developments in QAOA have shown that the optimized parameters tend to cluster around specific values for similar problem instances~\cite{Akshay_2021, brandao2018fixedcontrolparametersquantum}, suggesting that the optimized parameters obtained from simpler instances can serve as effective initialization for more complex ones~\cite{PhysRevA.104.L010401, brandao2018fixedcontrolparametersquantum, 10.3389/frqst.2023.1200975, galda2021transferabilityoptimalqaoaparameters, Monta_ez_Barrera_2025, Shaydulin_2023, Shi_2022, article123, montanez2025transfer, nguyen2025crossproblemparametertransferquantum, xu2025qaoaparametertransferabilitymaximum, sakai2024linearly}.


In Fig.~\ref{fig:Landscape_8-12}, parameterized solution landscape for $10n-u3R$ and $16n-u3R$ is generated using single layer QAOA ansatz. These plots exhibits comparable landscape in $(\gamma,\beta)$ plane, even though the cost values are different. The similarities of the landscapes provide insights that parameter transfer can be used from Fig.~\ref{fig:Landscape_8-12}(a)
to Fig.~\ref{fig:Landscape_8-12}(b). This claim is further strengthen by the clustering of optimized parameters represented in Fig.~\ref{fig:optimized_par} for higher depth ($p=15$) QAOA circuit. Here for any specific layer, optimized parameters of different nodes ($n_a=10, 12, .., 18$) of  $u3R$ graphs are concentrating at some specific value. This is the numerical proof that parameters of small instance problem can work as good initialization for more complex problem. 

The optimized parameters $\{\beta^{*}_i,\gamma^{*}_{i}\}$ of donor graph with nodes $n_d$ cluster for increasing $n_d$, we can consider transferring these optimized parameter as an initial guess for $\{\beta_i,\gamma_i$\} of the acceptor graph with nodes $n_a\geq n_d$ of the same family. In the next section, we have optimized targeted-single-layer to further improve the solution quality.

\subsection{Selective Optimization}



To find the targeted-single-layer $p_k\in[1,15]$ of acceptor graph with nodes $n_a\in[8,24]$ producing best $r_s$ (approximation ratio obtained by targeted-single layer optimization), we choose $50$ different graphs of fixed nodes ($n_a=8$) of a specific family and then selectively optimize each layer one-by-one to find the most probable layer that produces the highest optimal value, $r_s$.  Then we increase the number of nodes ($n_a=10, 12, .., 24$) and follow the same procedure. The complete procedure is given in  Algorithm~\ref{alg:1}.

\begin{algorithm}
\begin{algorithmic}[1]
\caption{Layer Selection Procedure for QAOA Parameter Transfer}
\label{alg:1}
\State \textbf{Initialize:} Set $p = 15$ layers in the QAOA circuit for an $n_d=8$ qubit U3R Hamiltonian. Initialize all $2p$ parameters using TQA method and optimize fully using Adagrad optimizer to obtain parameters $\{\gamma_1^*, \dots, \gamma_p^*, \beta_1^*, \dots, \beta_p^*\}$.
\For{$n_a \in \{8, 10, 12, 14, \dots, 24\}$}
    \State Transfer optimized parameters to acceptor $n_a$-qubit circuit, set $p = 15$.
    \State Initialize $A$ as a $p \times 9$ zero matrix.
    \For{experiment $= 1$ to $40$}
        \For{$p_k \in [1, p]$}
            \State Perform single-layer optimization for $p_k$.
            \State Compute resulting value $r_s$ for current $p_k$.
        \EndFor
        \State Identify $p_k^*$ corresponding to highest $r_s$.
        \State Increment $A[n_a, p_k^*]$ by 1.
    \EndFor
\EndFor
\State Divide each element of $A$ by $40$ to normalize.
\State \textbf{Output:} For each $n_a$, select most probable $p_k$ (layer) producing the highest $r_s$ from normalized $A$.
\end{algorithmic}
\end{algorithm}


\subsection{Regularization}
\begin{figure*}
	\centering
	\includegraphics[width=\linewidth]{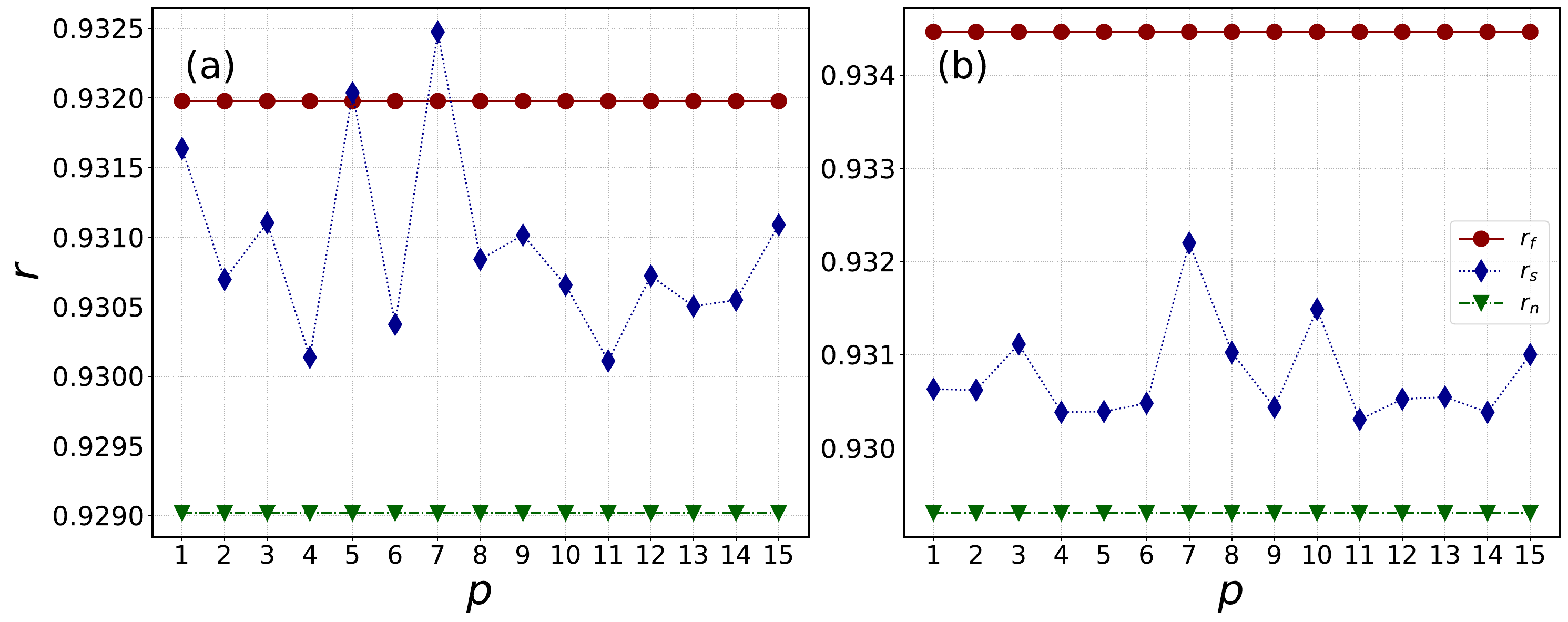}
	\caption{\textbf{$L_2$ Regularization:} Approximation ratios for $20n-u3R$ graph using various optimization techniques following parameters transfer initialization from $8n-u3R$ graph: (a) representing complex landscape trapping the full optimization, causing $r_s\geq r_f$, when any one of layers $5\ \rm{or}\ 7$ is selectively optimized. This problem is mitigated in (b) using $L_2$-Regularization which makes the parameter solution space smooth and improve the full optimization accuracy ($r_f$). Here $r_f,r_s$, and $r_n$ are approximation ratios of acceptor graph ($20n-u3R$), obtained by full, targeted-selective, and no optimizations respectively. Here regularization strength is taken as, $\lambda = 0.0001$}.
	\label{L2_reg}
\end{figure*}


\begin{figure*}
	\centering
	\begin{subfigure}[b]{0.49\textwidth}
    \caption{}
	\includegraphics[width=0.96\textwidth]{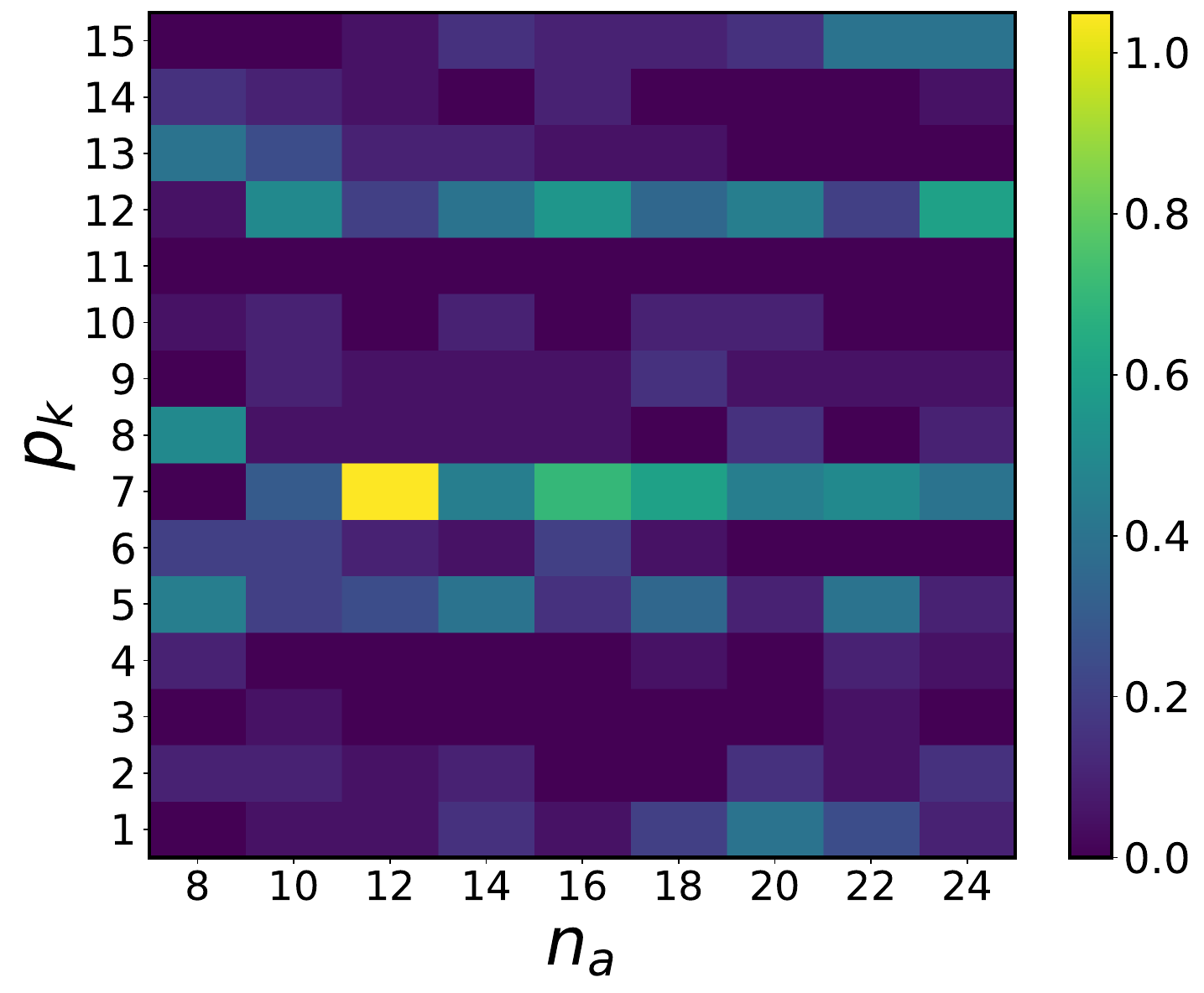}

	\label{fig:ls_u3R}
	\end{subfigure}
	\hfill
	\begin{subfigure}[b]{0.49\textwidth}
    \caption{}
	\includegraphics[width=\textwidth]{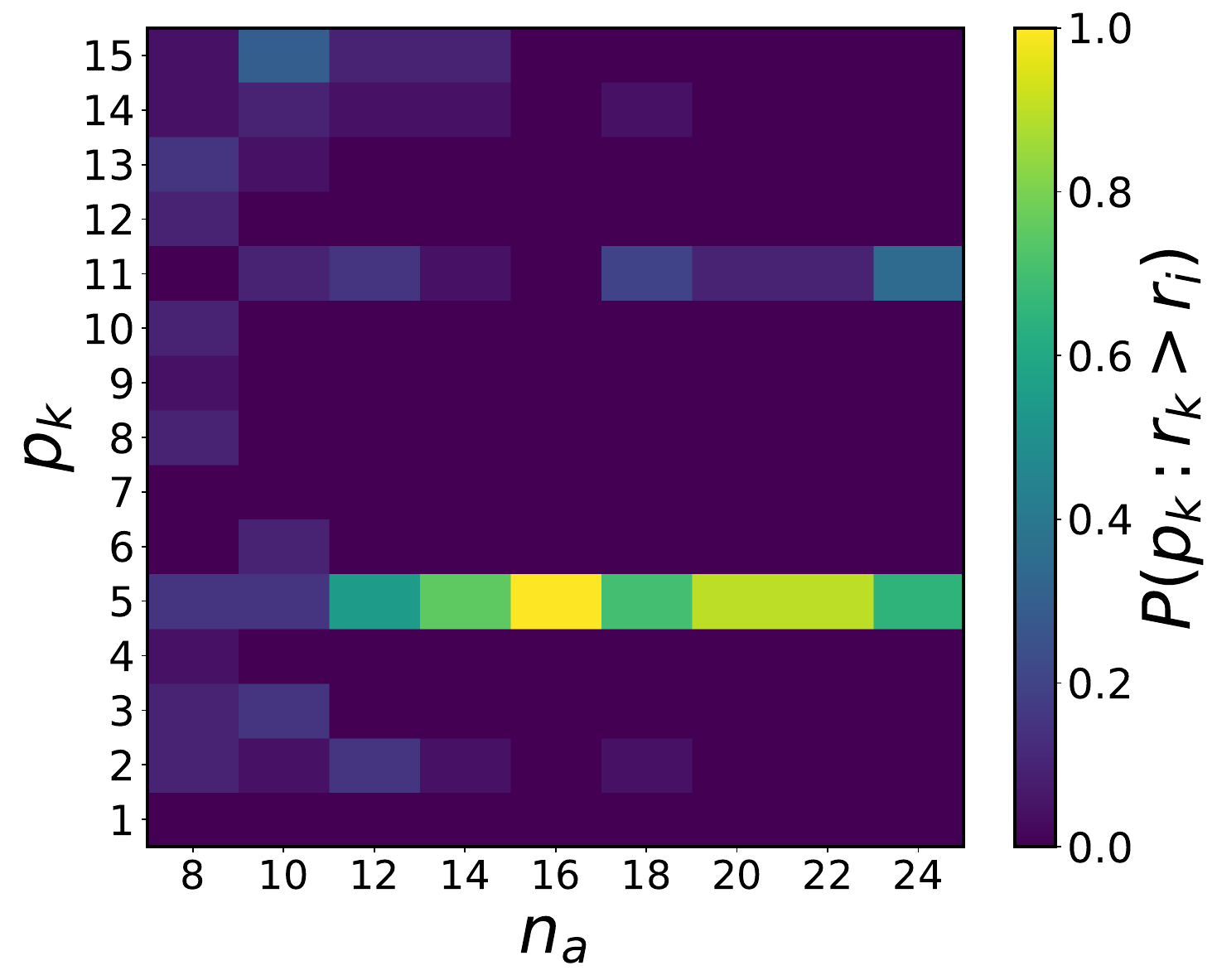}
	
	\label{fig:ls_uBA}
	\end{subfigure}
	\hfill
	\begin{subfigure}[b]{0.49\textwidth}
    \caption{}
	\includegraphics[width=0.96\textwidth]{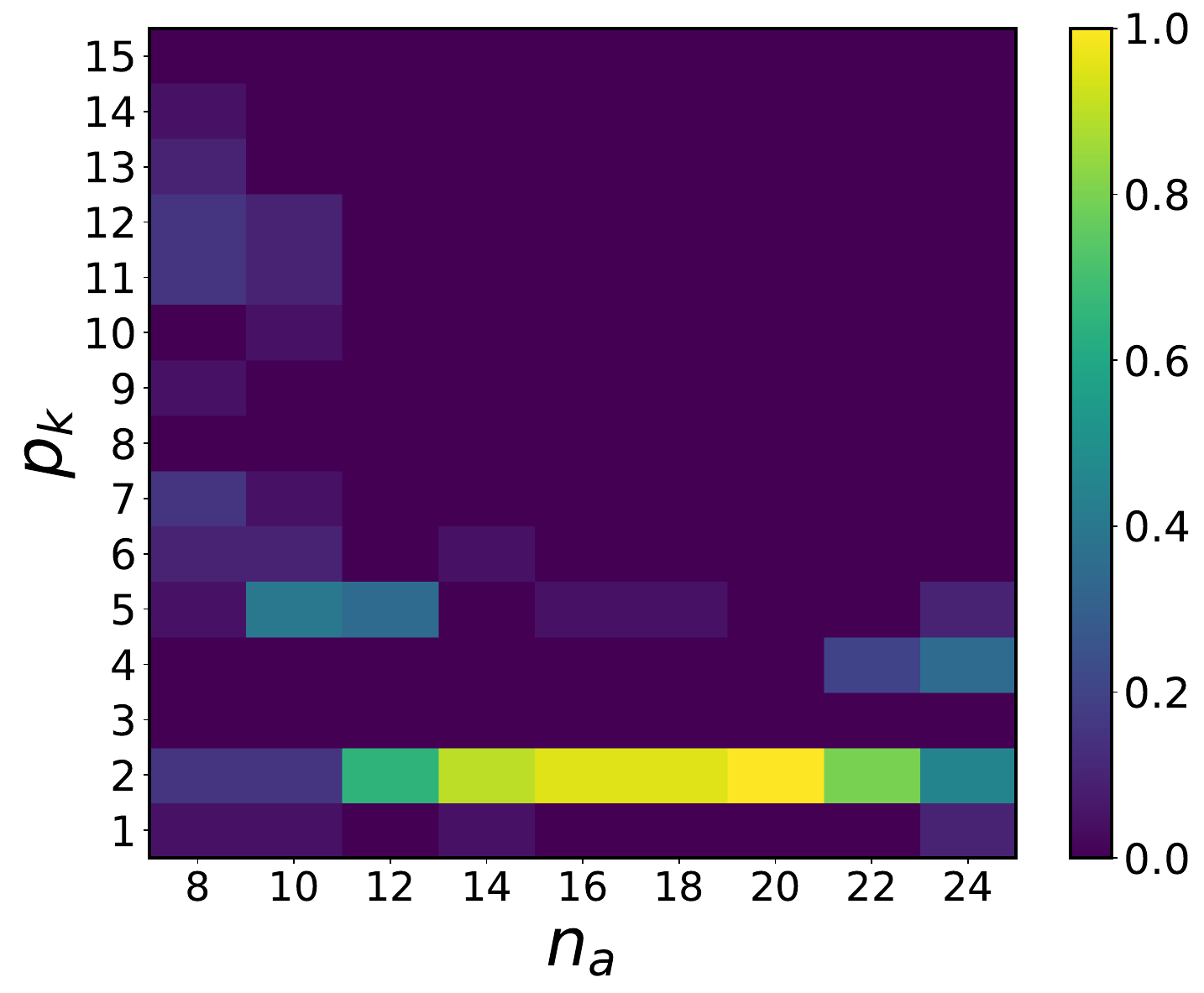}
	
	\label{fig:ls_uER}
    \end{subfigure}
	\hfill
	\begin{subfigure}[b]{0.49\textwidth}
    \caption{}
	\includegraphics[width=\textwidth]{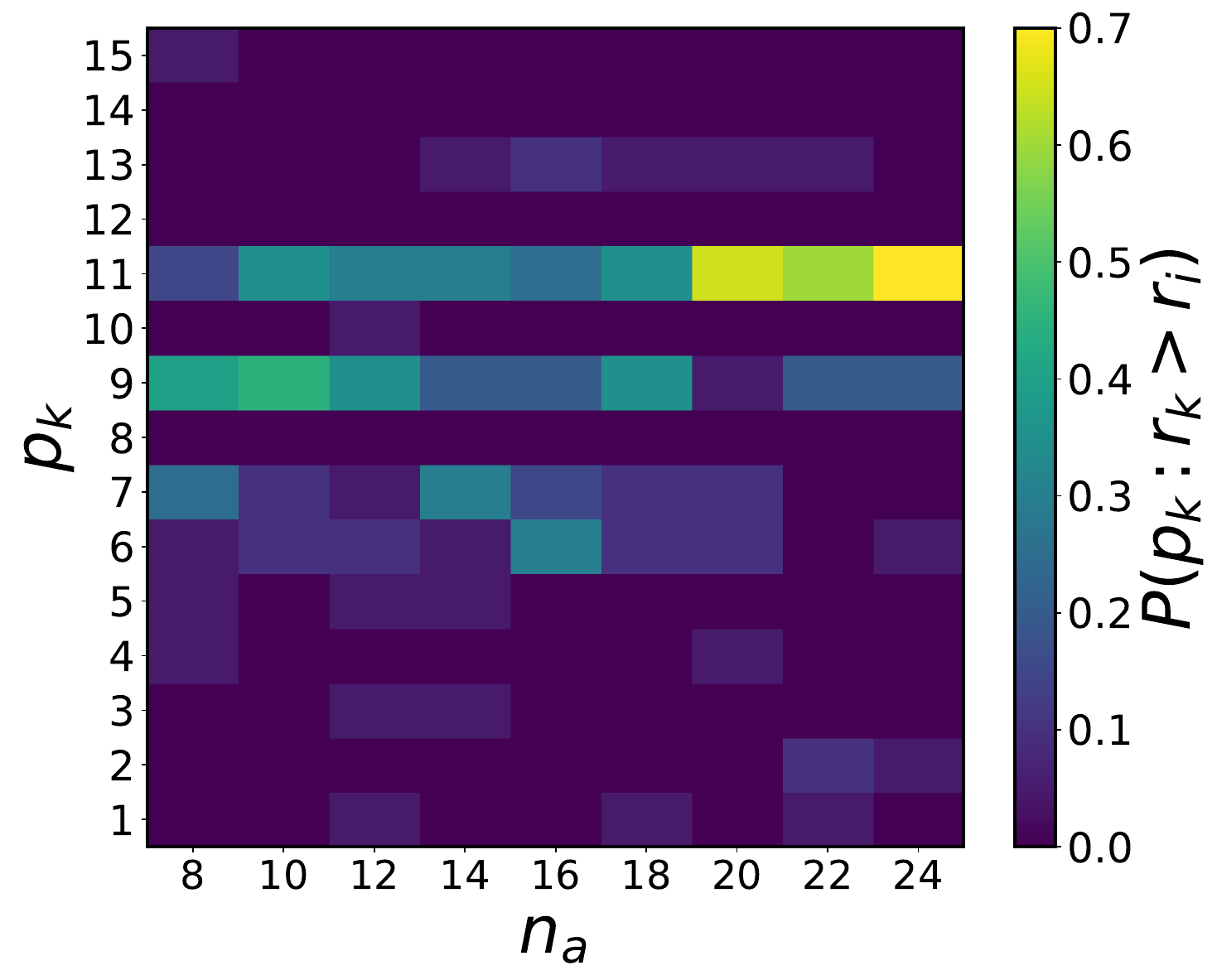}
	
	\label{fig:ls_w3R}
	\end{subfigure}
	\caption{\textbf{Optimal target layer:} Figure representing heat-maps to find the probability $P(p_k:r_k>r_i)$ of targeted layer $p_k$ providing best optimum cut ($r_k>r_i$) for (a) u3R, (b)Unweighted Barabási–Albert (uBA) graph, (c)Unweighted Erdős–Rényi (uER) graph, and (d) w3R graph families.}
	\label{fig:layer_selection}
\end{figure*}

As the depth ($p$) of QAOA ansatz increases, the number of parameters ($2p$) also increases, making the parameters solution space highly non-convex with many local optima. Thus, during full optimization, sometimes optimizer gets trapped in the local minima and consider it as global minima producing lower approximation ratio than selective optimization ($r_f \leq r_s$). One such incidence is presented in Fig.~\ref{L2_reg}(a). Here, 20n-u3R acceptor graph is optimized for both $2p$ and targeted-single-layer. 
During targeted-single-layer optimization ($r_s$) of any one of the layers $5,\ \rm{and}\ 7$, it outperformed the full optimization ($r_f$). We took $300$ sample cases to check the frequency of occurrence of these cases and found out that there were $12.78\%$ cases for u3R and $5.05\%$ cases for unweighted Erdős–Rényi (uER) graphs' instances respectively, where $r_s>r_f$.

This inconsistency can be mitigated by smoothing the parameterized solution space. For that purpose, many regularization techniques are available which reduces the complexity of the problem. We tested three regularization techniques for MaxCut problem and found out that $L_2$-regularization works best. Results are presented in Table~\ref{table:reg}.

\subsubsection{$L_1$ regularization (Lasso regularization)}
$L_1$ regularization \cite{schmidt2005least, lee2006efficient, lee2006efficient_in, 10.1111/j.1467-9868.2007.00607.x, wainwright2006inferring, goldstein2009split} also known as Lasso (least absolute shrinkage and selection operator) which adds the penalty term equal to the absolute sum of all parameters to the cost function. Mathematically of $L_1$ regularization is:
\begin{equation*}
		R_{L_1}(\gamma, \beta) = \sum_{i=1}^{p} (|\gamma_i| + |\beta_i|).
\end{equation*}

It is used to penalize the high value coefficients by reducing the coefficient value to zero, which helps in choosing the most important feature (variable) and removing less important ones. This aspect also allows lasso regression to handle some multicollinearity (high correlations among features) in a data set without affecting interpretability (without making the model hard to understand)~\cite{10.1111/j.2517-6161.1996.tb02080.x, ruppert2004elements}. However, lasso regularization is not suitable for high multicollinearity which makes it unsuitable for MaxCut.
	
\subsubsection{ $L_2$ regularization (Ridge regularization)}
Ridge regularization also known as $L_2$ regularization \cite{Hoerl01021970, tikhonov1943stability, cortes2012l2regularizationlearningkernels, 10.1093/gigascience/giaa133, patil2024optimalridgeregularizationoutofdistribution, vanlaarhoven2017l2regularizationversusbatch, https://doi.org/10.1002/jmri.24365, Hastie01102020, SACCOCCIO2014470} is most widely used regularization technique. It adds a penalty terms to the Ordinary Least Square (OLS) objective function that is proportional to the sum of squares of the parameters. It's mathematical form is:
	\begin{equation*}
		R_{L_2}(\gamma, \beta) = \sum_{i=1}^{p} (\gamma_i^2 + \beta_i^2).
	\end{equation*}
Ridge regularization shrinks parameters towards zero but not exactly zero. Compared to $L_1$ regularization which encourage sparsity by shrinking some parameters to zero, $L_2$ regularization aims to reduce the magnitude of all parameters while maintaining their importance. $L_2$ regularization also encourages smoother model output which may be beneficial when there is noise present in the system. It also effectively handles multicollinearity by distributing effects evenly.
	
\subsubsection{Smoothness Regularization}
	
	It enforces smoothness constraints in the parameters to control the complexity of a model~\cite{wahba1990spline, tikhonov1977solutions}. This adds a penalty term in the objective function that discourages abrupt changes in the function's evolution. It can be mathematically represented as:
	\begin{equation*}
		R_{sm}(\gamma, \beta) = \sum_{i=1}^{p-1} ((\gamma_{i+1} - \gamma_i)^2 + (\beta_{i+1} - \beta_i)^2).
	\end{equation*}
	It controls model's complexity by encouraging smooth transition of parameters. It is particularly effective in high-dimensional solution space and it reduces the over-fitting to noisy data.

\begin{table}[htb]
	\centering
	\begin{tabular}{|c|cccc|cccc|c|}
		\hline
		 & \multicolumn{8}{c|}{$n : \ (r_s>r_f)$} & \\
		\cline{2-9}
		$n_a$ & \multicolumn{4}{c|}{u3R} & \multicolumn{4}{c|}{uER} & N \\
		\cline{2-5} \cline{6-9}
		& $nr$ & $L_1$ & $L_2$ & $sm$ & $n_r$ & $L_1$ & $L_2$ & $sm$ & \\
		\hline
		08    & 57 & 32 & 22 & 36 & 25 & 23 & 15 & 20 & 300 \\
		10   & 42 & 36 & 24 & 31 & 30 & 26 & 15 & 17 & 300 \\
		12   & 22 & 18 & 08 & 18 & 15 & 04 & 02 & 02 & 300 \\
		14   & 46 & 31 & 16 & 18 & 09 & 07 & 01 & 08 & 300 \\
		16   & 35 & 15 & 11 & 04 & 07 & 02 & 03 & 09 & 300 \\
		18   & 28 & 23 & 18 & 22 & 05 & 03 & 02 & 03 & 300 \\
		\hline
	\end{tabular}
	\caption{Out of $N$ experiments, $n$ experiments producing $r_s>r_f$, where $r_s$ and $r_f$ are selective and full optimization approximation ratios respectively. Two graph families : unweighted 3 regular ($u3R$) and unweighted Erdős–Rényi ($uER$) are investigated. Here, $nr$: no regularization, $L_1$, $L_2$, $sm$ represent Lasso, Ridge and Smoothness regularization respectively.}
	\label{table:reg}
\end{table}

In Fig.~\ref{L2_reg}(b), $L_2$ regularization is used and it instantly fixes the problem by smoothing the solution space so that regularized full optimization achieves higher cost value than no regularization (Fig.~\ref{L2_reg}(a)), i.e, $r_f>r_s$. Regularization affects  $r_f, r_s,$ and $r_n$ as new parameters solution space is slightly different (smooth) than initial non-regularized one. So both, $r_n$ and 
$r_s$ are slightly changed but most significant improvement can be observed in $r_f$ which suffered from local optima trap without regularization due to large number ($2p$) of parameters optimization.

\begin{algorithm}
\caption{Parameters transfer and targeted-single-layer regularized optimization on QAOA  procedure}
\label{alg:transfer-targeted-fullopt}
\begin{algorithmic}[1]
  \Require Acceptor graph $G_a$ with $n_a \geq n_d$ nodes, QAOA depth $p = 15$
  \Require Transferred parameters $\{ \gamma^*_k, \beta^*_k \}_{k=1}^p$ from $8$-node donor graph (same family)
  \Require Optimal target layer index $k$ (from Algorithm~\ref{alg:1})
  \Require $L_2$ regularization strength $\lambda = 0.0001$
  \Require Cost Hamiltonian $H_C$, optimal classical value $C_{\max}$
  
  \Statex
  \State \textbf{(A) Parameter transfer initialization}
  \For{$k = 1$ to $p$}
    \State Set $\gamma_k \gets \gamma^*_k$
    \State Set $\beta_k \gets \beta^*_k$
  \EndFor

  \Statex

  \State \textbf{(B) Targeted-single-Layer \(L_2\)-regularized optimization}
  \State Fix all $[\gamma_j, \beta_j]$ for $j \neq k$
  \State Define cost function:
      $$
        \mathcal{L}_{\text{single}}(\gamma_{k}, \beta_{k}) = \bra{\psi} H_C \ket{\psi} + \lambda \left( (\gamma_{k})^2 + (\beta_{k})^2 \right)
      $$
  \State Use COBYLA to \textbf{optimize only} $(\gamma_{k}, \beta_{k})$ starting from $(\gamma^*_{k}, \beta^*_{k})$
  \State Let $(\gamma'_{k}, \beta'_{k})$ denote the optimal parameters
  \State Compute new quantum state with updated $\gamma'_{k}$, $\beta'_{k}$
  \State Compute $r_s \gets \dfrac{\bra{\psi'} H_C \ket{\psi'}}{C_{\max}}$

  \Statex

  \Statex
  \State \textbf{Output:} Return $r_s$
\end{algorithmic}
\end{algorithm}

\section{Results} \label{sec:results}
\subsection{Experimental Setup}
For donor graphs $n_d=8$,  we use  TQA ($t=0.75$)  algorithm for initialization of $2p$ parameters and then optimized all $2p$ parameters using Adagrad ($\rm{lr} = 0.1$, $\epsilon = 10^{-8}$, iterations$=100$) optimizer. These optimized parameters are taken as initial value for higher nodes ($n_a = 10, 12, ...., 24$) graphs of same family followed by targeted-single layer optimization using gradient free optimizer \enquote*{COBYLA}. The full procedure is presented in Algorithm~\ref{alg:transfer-targeted-fullopt}. For efficiency plots (Fig.~\ref{fig:eff_acc_u3R}, Fig.~\ref{fig:eff_acc_uBA}..., Fig.~\ref{fig:eff_acc_wER}), 40 graphs of each nodes are simulated for a specific family and then mean and standard deviation are calculated to plot the approximation ratio and optimization time with error-bars. The computational resource details are provided in Appendix~\ref{app:appB}.
\subsection{Targeted-Single Layer optimization}
Here we discuss about the optimization of the acceptor graphs $n_a\in[8,24]$. We have taken  graphs from u3R,unweighted Barabási–Albert, unweighted Erdős–Rényi, and w3R families.

In Fig.~\ref{fig:layer_selection}(a-d) we plot the heat map of $r_s$ corresponding to optimized individual layer $p\in[1,15]$ for each node $n_a\in[8,24]$ of the acceptor graph.  For example, for $n_a=8$, u3R graph, we optimized all layers individually $p\in[1,15]$ and found that layer $p=8$ is most likely to produce best $r_s$ as shown in Fig.~\ref{fig:layer_selection}(a). Thus, from Figs.~\ref{fig:layer_selection}(a-d), we found that layers $p=7, 5, 2$ and $11$ are most probable targeted-single layers capable of producing highest $r_s$ during single layer optimization for u3R, unweighted Barabási–Albert, unweighted Erdős–Rényi, and w3R,  respectively.

This technique significantly reduces the optimization time at the cost of slight reduction of approximation ratio as only 2 parameters are optimized. Thus, there is a trade-off between efficiency and accuracy of QAOA, but the improvement in the efficiency is very large as compared to the reduction in accuracy. We compared the value of $r$ for different families of graphs using different optimizers and reported in Table \ref{table:r_trend}. From the table, we can verify that, for all the considered graph family:
\begin{equation}
	r_{n} \leq r_{s} \leq r_{f}.
	\label{r_trend}
\end{equation}


\begin{table*}[htbp]
	\centering
	\begin{tabular}{|c|ccc|ccc|ccc|ccc|ccc|}
		\hline
		$n_a$ & \multicolumn{3}{c|}{u3R} & \multicolumn{3}{c|}{uBA} & \multicolumn{3}{c|}{uER} & \multicolumn{3}{c|}{w3R} & \multicolumn{3}{c|}{wER}\\
		\cline{2-16}
		& $r_n$ & $r_s$ & $r_f$ & $r_n$ & $r_s$ & $r_f$ & $r_n$ & $r_s$ & $r_f$ & $r_n$ & $r_s$ & $r_f$ & $r_n$ & $r_s$ & $r_f$\\
		\hline
		08 & 0.98220 & 0.98414 & 0.98997 & 0.99816 & 0.99826 & 0.99826 & 0.97791 & 0.97955 & 0.98119 & 0.95823 & 0.96297 & 0.97366 & 0.94754 & 0.95194 & 0.96861 \\
		10 & 0.96117 & 0.96389 & 0.96869 & 0.97121 & 0.97249 & 0.98261 & 0.96950 & 0.97096 & 0.97959 & 0.95445 & 0.95760 & 0.96635 & 0.92018 & 0.92468 & 0.95524 \\
		12 & 0.95807 & 0.96055 & 0.96425 & 0.95026 & 0.95175 & 0.96658 & 0.94402 & 0.94752 & 0.95939 & 0.95225 & 0.95523 & 0.96220 & 0.89903 & 0.90447 & 0.94548 \\
		14 & 0.95598 & 0.95817 & 0.96072 & 0.94158 & 0.94329 & 0.95926 & 0.93694 & 0.93942 & 0.95582 & 0.95234 & 0.95474 & 0.96151 & 0.86670 & 0.87544 & 0.93167 \\
		16 & 0.94696 & 0.94947 & 0.95273 & 0.92994 & 0.93216 & 0.94933 & 0.92855 & 0.93110 & 0.94502 & 0.94741 & 0.94941 & 0.95603 & 0.84191 & 0.85236 & 0.91880 \\
		18 & 0.95155 & 0.95339 & 0.95607 & 0.92498 & 0.92689 & 0.94762 & 0.88888 & 0.89418 & 0.92646 & 0.94245 & 0.94449 & 0.94971 & 0.82542 & 0.83586 & 0.90864 \\
		20 & 0.94861 & 0.95023 & 0.95282 & 0.91745 & 0.91879 & 0.94131 & 0.88304 & 0.89869 & 0.93017 & 0.94505 & 0.94724 & 0.95205 & 0.81365 & 0.82472 & 0.90190 \\
		22 & 0.94481 & 0.94633 & 0.94846 & 0.91095 & 0.91200 & 0.93727 & 0.87578 & 0.88851 & 0.92404 & 0.94090 & 0.94283 & 0.94781 & 0.80325 & 0.81336 & 0.88830 \\
		24 & 0.94897 & 0.95059 & 0.95295 & 0.90445 & 0.90533 & 0.93266 & 0.86540 & 0.87972 & 0.92040 & 0.93899 & 0.94059 & 0.94566 & 0.79751 & 0.80753 & 0.87708 \\
		\hline
	\end{tabular}
	\caption{\textbf{Comparison of $r$ across different families:} Approximation ratio, $r$ using various optimization techniques for different nodes graphs of unweighted ($u$) families: 3-regular ($u3R$), Barabási–Albert ($uBA$) and Erdős–Rényi ($uER$) graphs and also weighted families: 3-regular ($w3R$) and Erdős–Rényi ($wER$) graphs. Here, $r_n$ : no optimization, $r_s$ : targeted-single layer optimization and $r_f$: full optimization after parameter transfer.}
	\label{table:r_trend}
\end{table*}


\subsection{Efficiency in unweighted graph families }

\begin{figure*}
		\includegraphics[width=\textwidth]{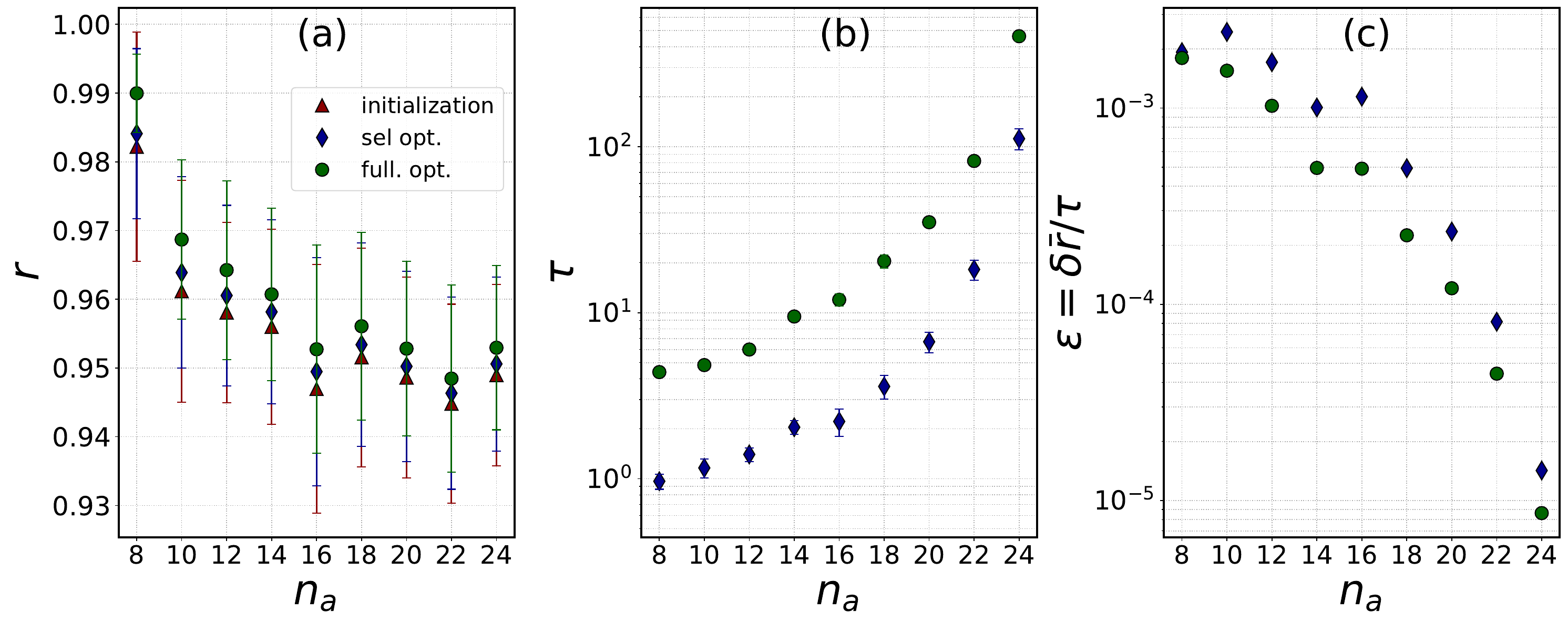}
		\caption{\textbf{Parameters Transfer followed by targeted-single-layer optimization in u3R:} Plots for (a) approximation ratio, (b) optimization time  and (c) efficiency. For initialization, $n_d=8$ nodes graphs are used as donor. The filled circle is for full optimization and the corresponding approximation ratio as $r_f$, the filled diamond are for selective-targeted-layer optimization with $r_s$ as the corresponding approximation ratio, and filled upper triangle for initial value of $r$ denoted as $r_n$. }
        \label{fig:eff_acc_u3R}
    \end{figure*}
	
\begin{figure*}
		\includegraphics[width=\textwidth]{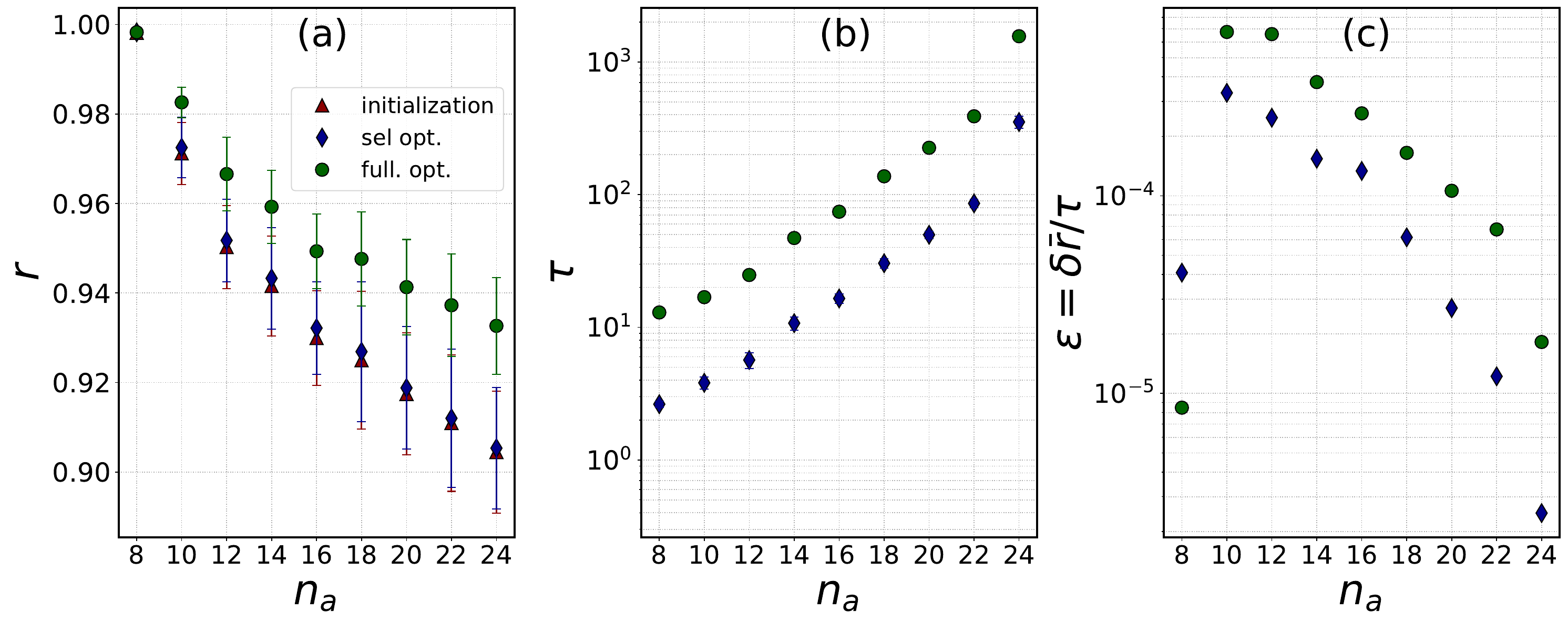}
		\caption{Same as in Fig.~\ref{fig:eff_acc_u3R} for uBA graph}
	\label{fig:eff_acc_uBA}
\end{figure*}
	
\begin{figure*}
		\includegraphics[width=\textwidth]{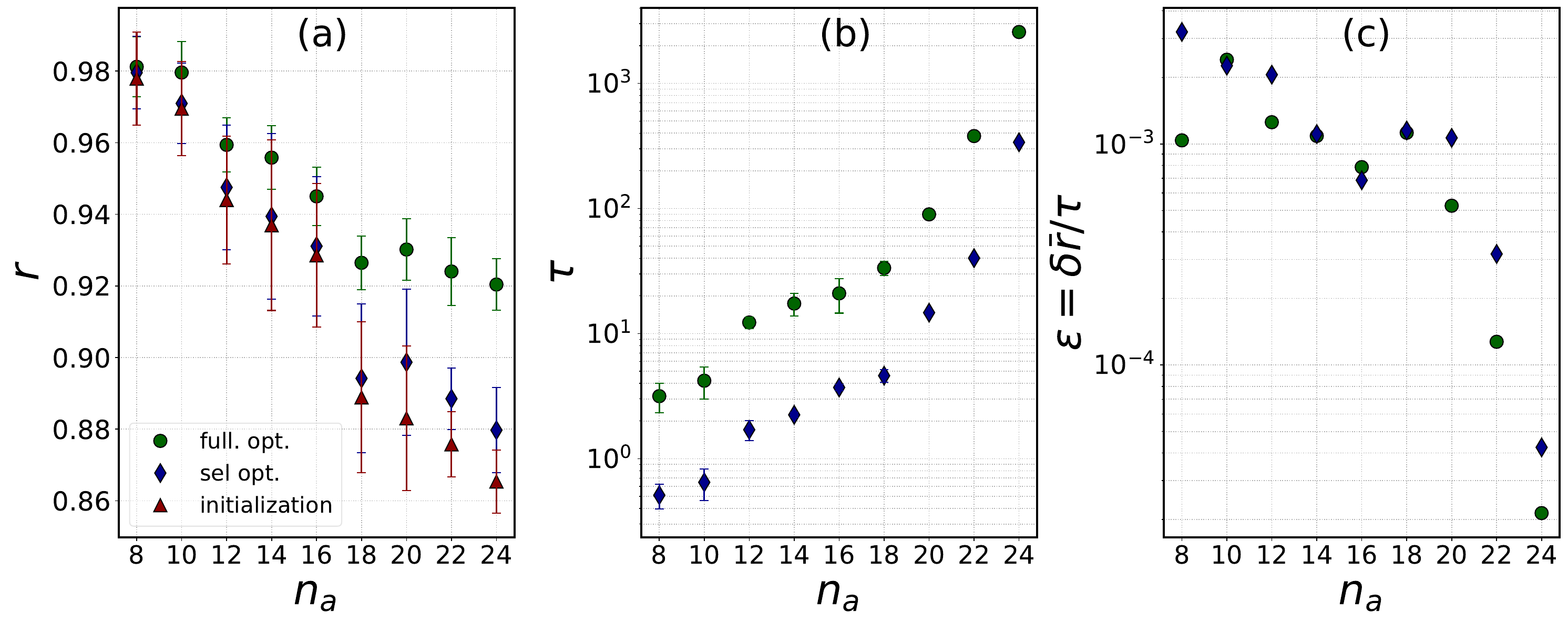}
		\caption{Same as in Fig.~\ref{fig:eff_acc_u3R} for  uER graph}
		\label{fig:eff_acc_uER}
\end{figure*}

Now, we consider unweighted graph families (u3R, Erdős–Rényi (uER), and Barabási–Albert (uBA)) and test the QAOA, integrated with parameter transfer followed by targeted-single layer optimization using $n_d=8$ as donor and $n_a\in[8,24]$ as acceptor graphs. To compare the efficiency of targeted-single-layer optimization with that of full and no optimization, we consider total time $\tau$ taken by the given optimizer for a given acceptor graph with nodes $n_a$ and the ratio of difference between optimized value of cost function with that of no optimization  and $\tau$.

We find the optimized targeted single layer for each family using Figs.~\ref{fig:layer_selection}(a-c). 
Employing targeted-single-layer $p_{k}=7$ for u3R, we plot $r$, optimization time ($\tau$), and efficiency ($\epsilon = \delta \bar{r}/\tau$) in Fig.~\ref{fig:eff_acc_u3R}(a-c), respectively. From Fig.~\ref{fig:eff_acc_u3R}(a), one can observe that the  initialization without optimization (shown by green upper triangle) is producing  $r_{n} > 0.878$, which is further improved to $r_s$ by optimizing only targeted-single-layer $p_{k}=7$. Also, $r_{n}$ is very close to the $r_{f}$ (shown by filled red circle), indicating nearly optimal initialization which also verifies the transferability of optimized parameters between similar instances. In targeted single layer only $2$ parameters require optimization, thus, $\tau_{s} < \tau_{f}$ as can be seen from Fig.~\ref{fig:eff_acc_u3R}(b). Here $\tau_s$ and $\tau_f$ are optimization time for targeted-single and full layers, respectively. It is observed that $\tau_{f} \approx 8\times \tau_{s}$. We further plot efficiency, defined as $\epsilon=\frac{r_n-r}{\tau}$, in Fig.~\ref{fig:eff_acc_u3R}(c). It is clear from the plots that the numerator is almost constant,  hence the efficiency is inversely proportional to $\tau$. Thus, the selective optimization is more efficient than the full optimization for all nodes $n_a\in[8,24]$.

Similarly, we investigated MaxCut in unweighted Barabási–Albert (uBA) family of graphs  (with edge probability $6$ for each nodes graph) and show the results in in Fig.~\ref{fig:eff_acc_uBA}(a-c).  The qualitative behavior of this family of graph is similar as in the case of unweighted u3R family. Here, initialization gives slightly reduced  value of $r_n$ as compared to u3R, which can be due to the higher connectivity in this family. However, we still achieved accuracy better than the classical GW algorithm's threshold ($r_{GW}\geq 0.878$) which is further improved by targeted-single-layer optimization of  $p_k=5$ which was achieved by Fig.~\ref{fig:ls_uBA}. From Fig.~\ref{fig:eff_acc_uBA}(a), it is clear that as the number of nodes $n_a$ increasing, the value of $r_s$, from selective layer optimization  decreasing almost uniformly, yet still it is better than the classical value. The trend of optimization time $\tau$ and efficiency $\epsilon$ are shown in Fig.~\ref{fig:eff_acc_uBA}(b) and Fig.~\ref{fig:eff_acc_uBA}(c), respectively. These plots further conforms that the selective layer optimization is more efficient.


Now, more complex graph family, unweighted Erdős–Rényi (uER) is investigated in Fig.~\ref{fig:eff_acc_uER}, where initialization only, is unable  to produce $r_{n} \geq r_{GW}$ for higher nodes graphs ($n_a\geq 22$), as presented in Fig.~\ref{fig:eff_acc_uER}(a). In those cases, targeted-single layer optimization helps us to achieve $r_{s}\geq r_{GW}$, as one can also see in Table \ref{table:r_trend}. 
In Table \ref{table:r_trend}, one can observe that for less nodes ($n\leq18$) graphs in uER, ($r_f-r_s$) is relatively low as compared to higher nodes graphs resulting only slight improvement in the efficiency ($\epsilon_s-\epsilon_f$) for less nodes graphs, but very noticeable improvement can be observed in higher nodes ($n > 18$) graphs presented in Fig.~\ref{fig:eff_acc_uER}(c).

\subsection*{Efficiency in weighted graph families}
\begin{figure*}[htb]
		\includegraphics[width=\textwidth]{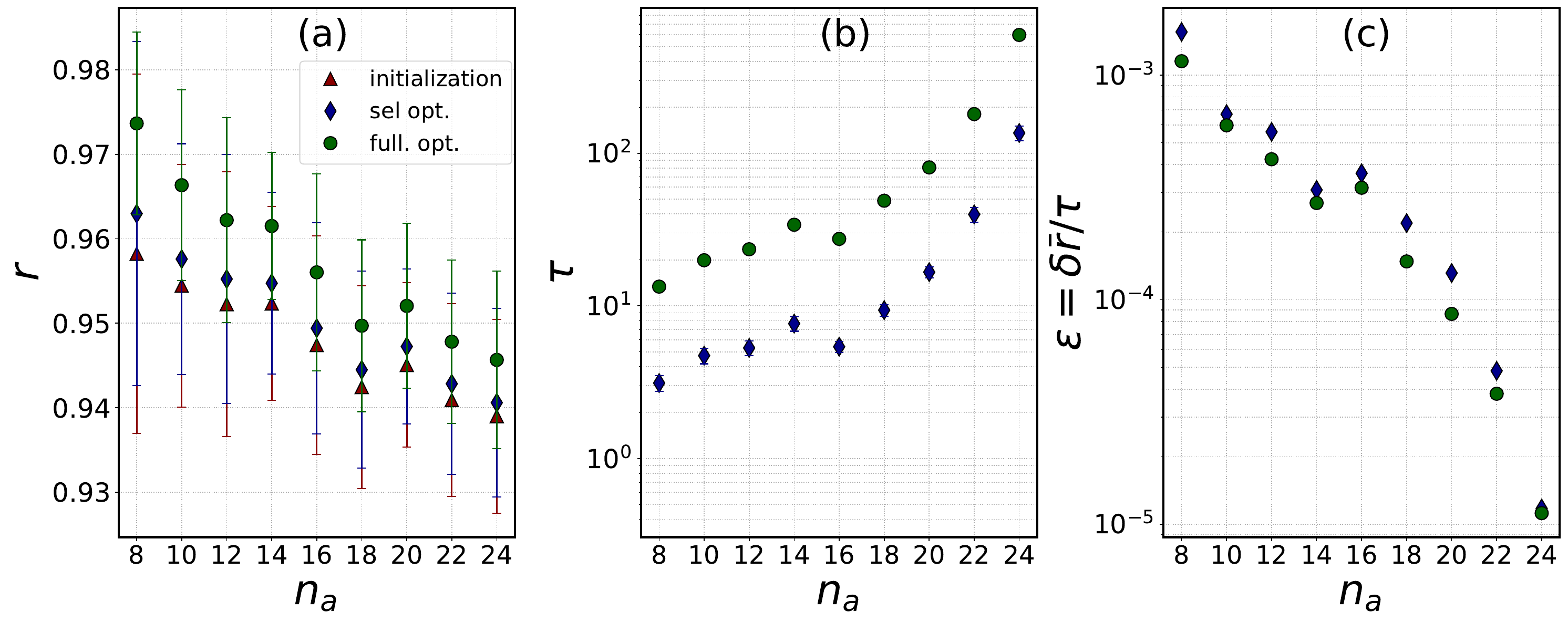}
		\caption{Same as in Fig.~\ref{fig:eff_acc_u3R} for w3R.}
		\label{fig:eff_acc_w3R}
\end{figure*}
\begin{figure*}[htb]
	\includegraphics[width=\textwidth]{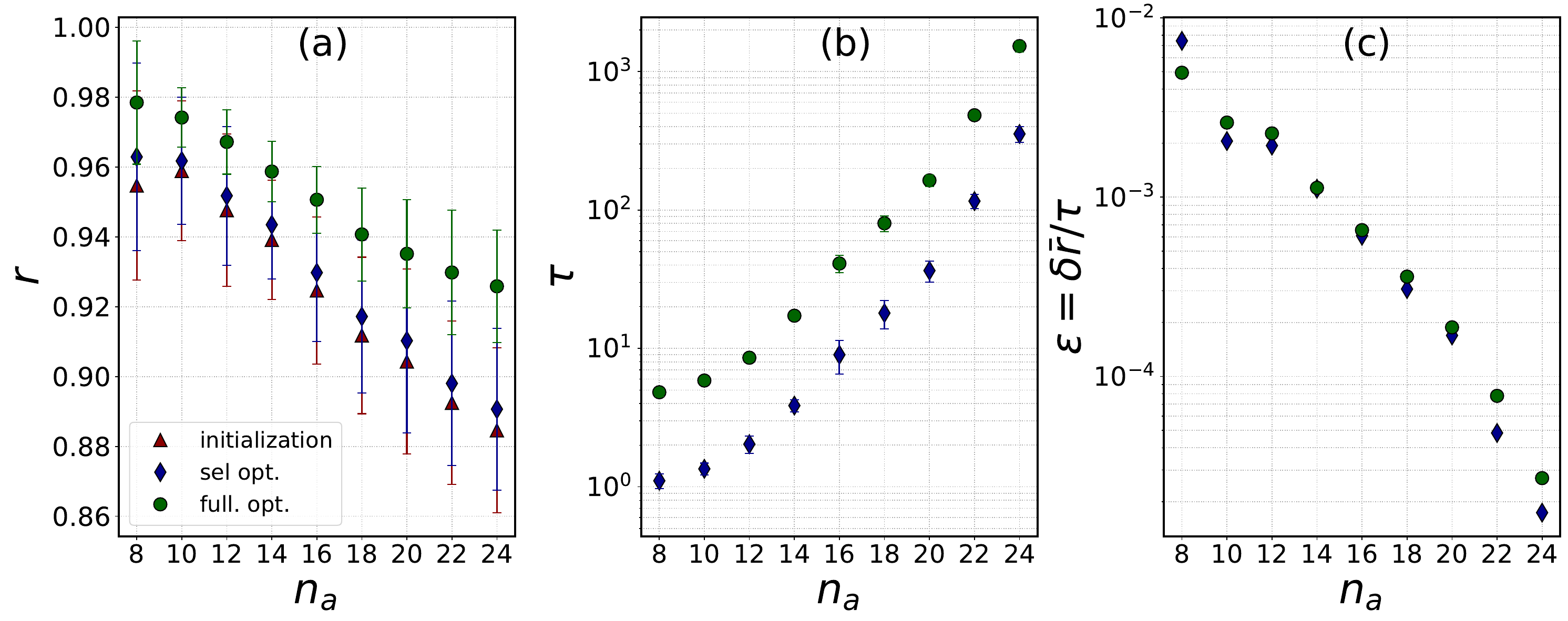}
	\caption{Same as in Fig.~\ref{fig:eff_acc_u3R} for wBA graph.}
	\label{fig:eff_acc_wBA}
\end{figure*}
\begin{figure*}[htb]
	\includegraphics[width=\textwidth]{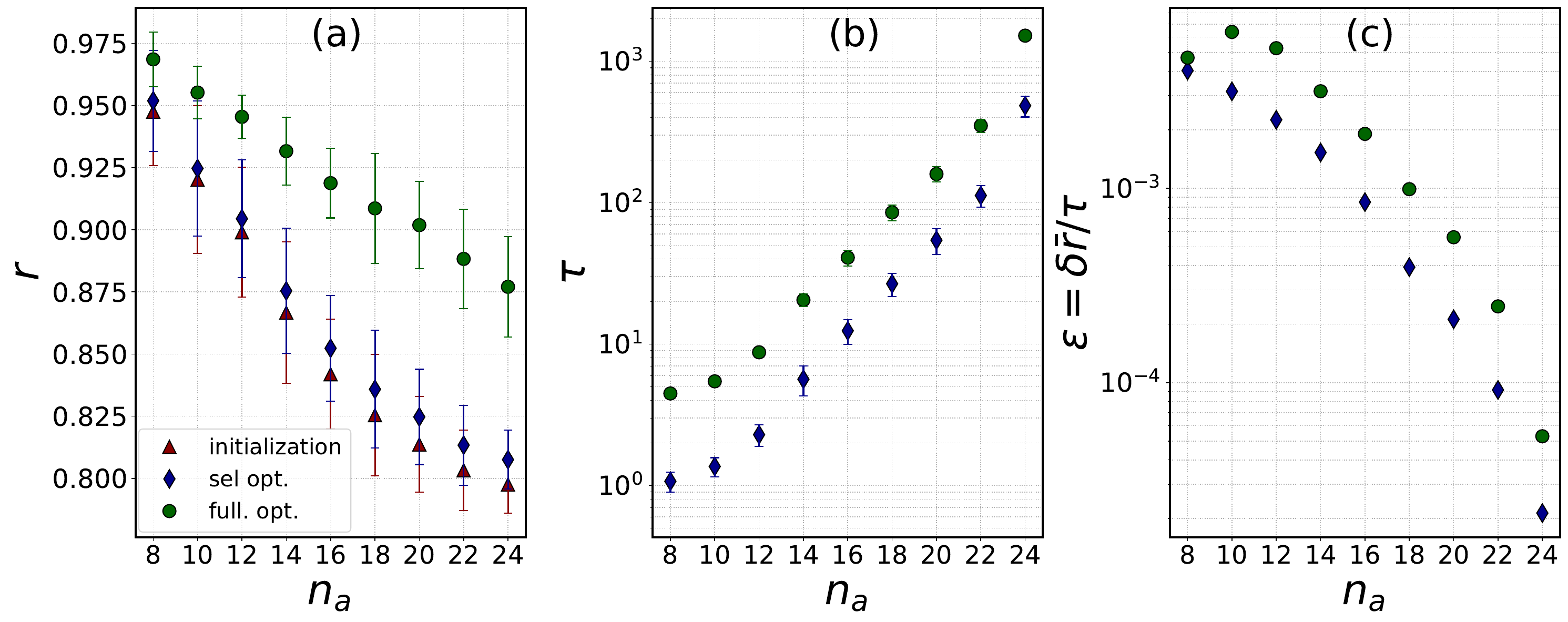}
	\caption{Same as in Fig.~\ref{fig:eff_acc_u3R} for wER graph.}
	\label{fig:eff_acc_wER}
\end{figure*}

Weighted graphs are more complex, as two fixed node graphs of same family can have different MaxCut (maximum weight), making them completely different from each other. We choose all three weighted families (w3R, wBA and wER) using Gaussian edge-weights ($\mu$ = $1.0$, $\sigma$ = $0.5$).

The w3R family achieves initialization $r_n>r_{GW}$ for all nodes as shown in Fig.~\ref{fig:eff_acc_w3R}(a). Here we used $n_d=8$ node graph of w3R for parameter transfer to $n_a\in[8,24]$ nodes acceptor graphs of w3R family. The value of  $r_n$ is further improved to $r_s$ by targeted-single layer optimization of layer $p_k=11$  (obtained from Fig.~\ref{fig:ls_w3R}). It is to be noted that for any  $n_a$ the value of $r_s$ is higher for u3R as compared to that of w3R. Also, optimization time ($\tau$) in Fig.~\ref{fig:eff_acc_w3R}(b) is approximately equal to the unweighted case in Fig.~\ref{fig:eff_acc_u3R}(b). Thus, we conclude that selective optimization is more efficient, as inferred from Fig.~\ref{fig:eff_acc_u3R}(c). This suggests, we can use parameters transfer technique followed by targeted-selective layer optimization in w3R.

Now, weighted Erdős–Rényi (wER) family with Gaussian weights is investigated using the selective layer $p_k=2$ and results are plotted in Fig.~\ref{fig:eff_acc_wER}. Here we used $n_d=8$ node graph of wER family for parameter transfer to $n_a\in[8,24]$ nodes acceptor graphs.
One can observe that for small instances ($n_a\leq 12$) initialization gives $r_{n}\geq 0.878$, but for larger  instances ($n_a\geq14$), $r_{n}$ starts decreasing with the increasing $n_a$ and at $n_a=22,24$ we get $r_{n}\approx 0.80$, which is far from the $r_{GW}\geq0.878$). We performed targeted-single-layer optimization using $p_k=2$ and found that $r_n$ can be improved to $r_s$. We observed that  $r_s$  does not reach to $r_{GW}$ for complex (higher nodes) instances as shown in Fig.~\ref{fig:eff_acc_wER}(a). So selective optimization is not efficient in wER as shown in Fig.\ref{fig:eff_acc_wER}(c). In~\cite{cjjm-87gl}, they used random edge weights and showed that this approach can not be used in wER family. Thus, full optimization is required in wER after parameters transfer initialization to achieve $r\geq r_{GW}$.

The complexity of wBA family is hybrid of w3R and wER. Thus, we have achieved approximation ratios, $\{r_n, r_s,r_f\}_{wER}\leq \{r_n, r_s,r_f\}_{wBA}\leq \{r_n, r_s,r_f\}_{w3R}$. The results for wBA approximation ratios are plotted in Fig.~\ref{fig:eff_acc_wBA}(a). Optimization time ($\tau$) is plotted in Fig.~\ref{fig:eff_acc_wBA}(b), which gives $\tau_f \approx 5\times \tau_f$. One can observe from Fig.~\ref{fig:eff_acc_wBA}(c) that efficiency full optimization is dominating slightly $\forall \ n_a\geq10$. This conclude that targeted single layer optimization is not efficient in the wBA family.

\subsection{Effect of different initialization}
We used three different methods to initialize parameters of the acceptor graphs. These are parameters transfer, TQA, and randomly generated values of $\gamma_i$ and $\beta_i$.
\begin{figure*}[htb]
	\centering
	\includegraphics[width=\linewidth]{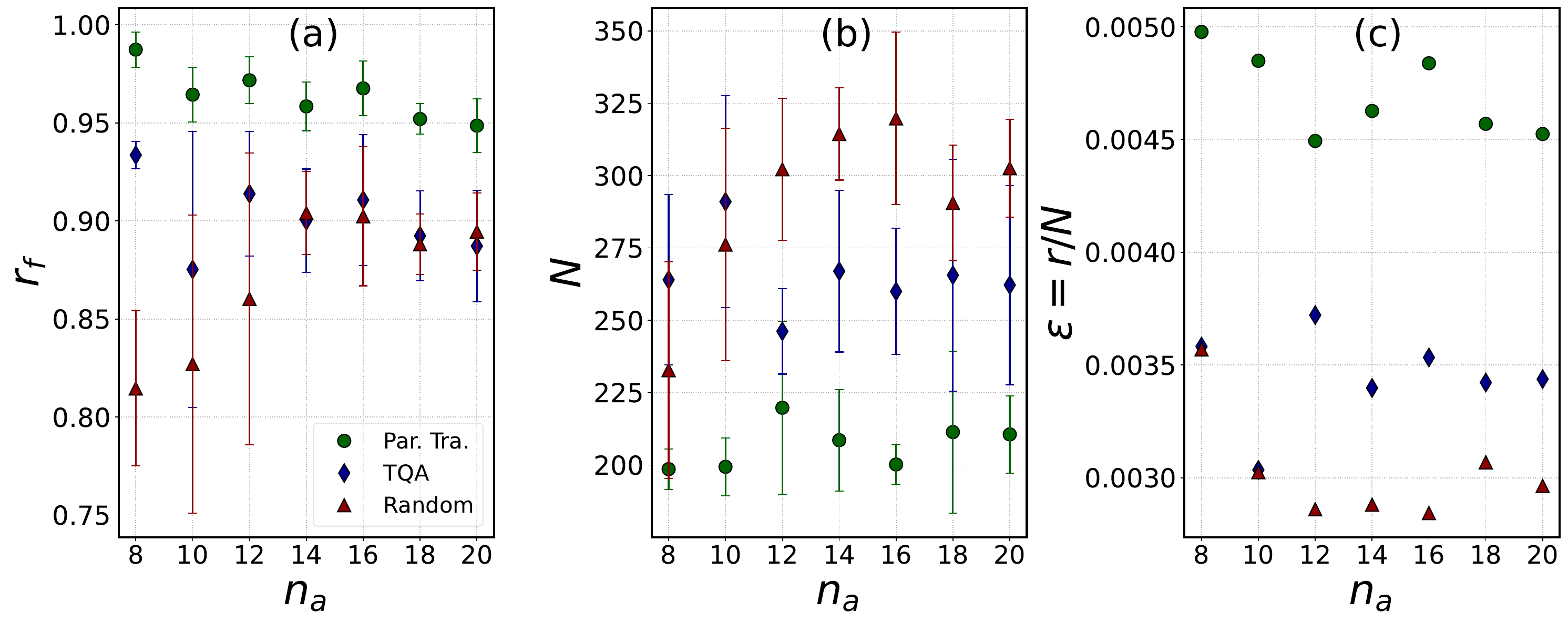}
	\caption{\textbf{Different initialization for u3R:} Plots for (a) approximation ratio, (b) required optimization steps (N), and (c)efficiency is compared for u3R using different initialization. The filled circles are for parameter transfer, the filled diamonds are for TQA, and filled upper triangles are for  parameters generated randomly and uniformly between $0$ to $\pi$.}
	\label{initialization_u3R}
\end{figure*}

For all the initialization, full optimization results of $r_f$ are shown in Fig.~\ref{initialization_u3R}. In Fig.~\ref{initialization_u3R}(a) average $r_{f}$ for different nodes and initialization is plotted and one can observe that parameters transfer initialization always outperformed others. This can be attributed to the fact that during parameter transfer, the initialization is nearly optimal for same family of graphs. Thus, there is very less chances of optimizer getting stuck in local optima and it find global optima easily. But in the case of other initialization, the starting point is far away from global optima and there are many local optima in between, which can trap the optimizers to produce false global optima. In Fig.~\ref{initialization_u3R}(b), optimization steps are plotted for each initialization and we can see that parameter transfer always requires least optimization steps. Thus,  it is clear that parameters transfer is the most efficient technique to achieve highest approximation ratio in least optimization steps than any other initialization as can be seen from  Fig.~\ref{initialization_u3R}(c).

\subsection{Effect of different optimizer}
\begin{figure*}[htb]
	\centering
	\includegraphics[width=\linewidth]{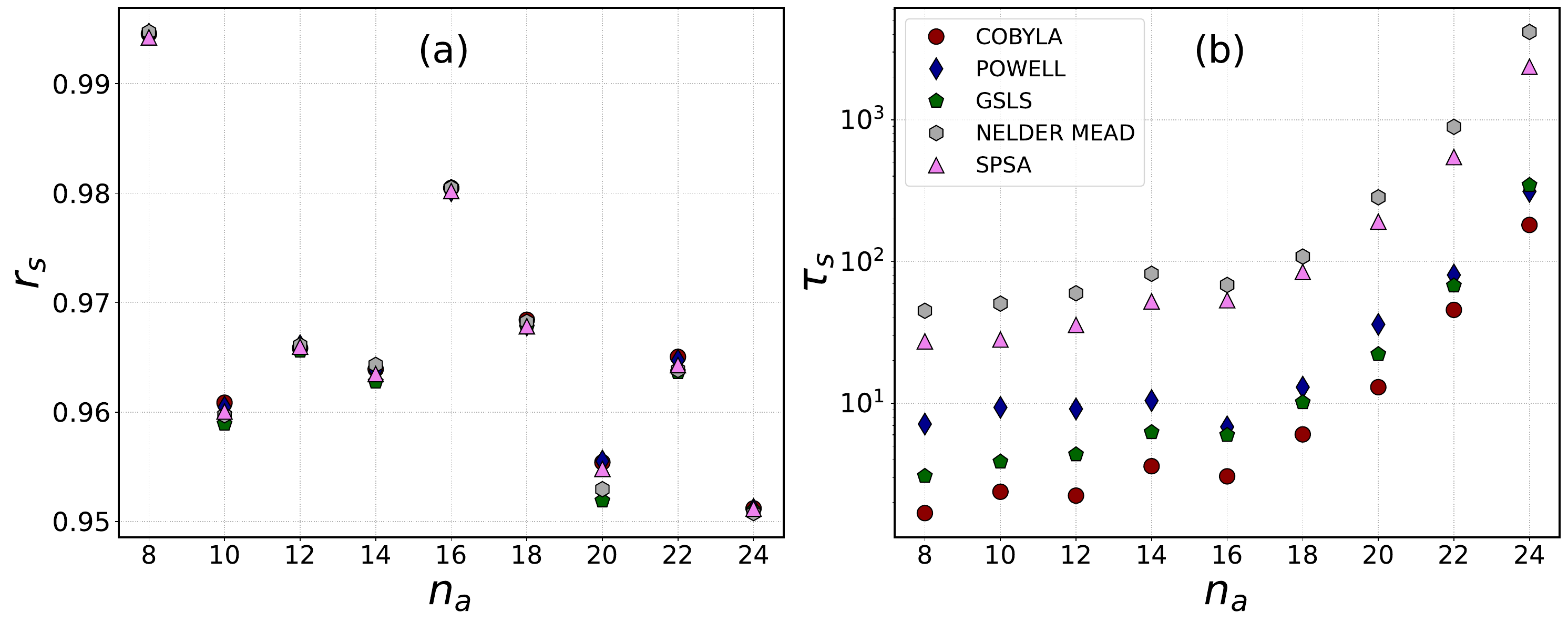}
	\caption{{\bf Effect of different optimizers:} Targeted-single layer optimization of $u3R$ family using various optimizers. Points in (a) represent $r_s$ vs $n_a$ for different optimizers and it can be observed that all the investigated optimizers are producing approximately same $r_s$, suggesting no difference in terms of accuracy, but in (b) optimization times ($\tau_s$) are different, thus, \enquote*{COBYLA} is most reliable out of all the investigated optimizers.}
	\label{optimizers_u3R}
\end{figure*}
Now, we are going to test other optimizers (POWELL, GSLS (Gaussian-smoothed Line Search), NELDER-MEAD, SPSA (Simultaneous Perturbation Stochastic Approximation)) along with COBYLA (Constrained Optimization By Linear Approximation) for selective optimization. The results are plotted in Fig.~\ref{optimizers_u3R}(a) for different nodes ($n_a$) of u3R graphs. Here, 40 graphs of each nodes are taken to plot the average approximation ratio ($r_s$) for parameters transfer followed by targeted-single-layer optimization with $p_k=7$ using different optimizers. Here, in Fig.~\ref{optimizers_u3R}(a), one can observe that there is no significant difference in the $r_{s}$ for various optimizers. However, optimization time of these optimizers are different as can be observed in Fig.~\ref{optimizers_u3R}(b), here, $\tau_s$ is selective optimization time of different optimizers. So in case of accuracy all the optimizers can work well but considering optimization time along with accuracy, the optimizer with least optimization time is more suitable. Thus, COBYLA optimizer is most appropriate out of all other optimizers investigated.

\section{Conclusion} \label{sec:conclusion}
The cost function landscape of the MaxCut problem, when optimized using QAOA, is highly non-convex. Consequently, finding globally optimal parameters becomes exponentially hard with the system size 
$n$ as the optimization implicitly explores the solution landscape defined over a $2^n$-dimensional Hilbert space. To compensate this, we report parameter transfer initialization followed by targeted-single-layer optimization strategy in QAOA to obtain approximate solutions for the MaxCut problem using numerical simulation of various graph families: u3R, uBA, uER, w3R, wBA and wER.

Due to high number of parameters and complex graphs, sometimes optimizer stuck in local optima during full optimization causing $r_s\geq r_f$.  This inconsistency is mitigated using $L_2$ (ridge) regularization that makes the parameters solution space smooth so that optimizer can avoid local optima trap during full optimization. There are other regularization but we found that $L_2$ works best in MaxCut. All the results presented in the the main text are regularized using $L_2$.

As only two parameters are being optimized in targeted single-layer optimization of acceptor graph,  we need very good initialization. We start with a small problem ($n_d = 8$) and optimize all $2p$ parameters and use the technique of transferability of parameters between similar instances. These optimized parameters are used as initialization for complex problem ($n_a\geq n_d$). This initialization produces the nearly optimal solution (i.e., $r_n \approx r_f$). 

To further improve the cost value, we use targeted-single-layer ($p_k$) optimization of acceptor graphs ($n_a$) after parameters transfer initialization. Yet, the problem occurs about the choice of targeted-single-layer. We found out that for a specific family, there is a target-layer which produces better $r_{s}$ than others, when selectively optimized. Using $p= 15$ layers of QAOA ansatz and COBYLA optimizer, for unweighted 3-Regular (u3R), Erdős–Rényi (uER), Barabási–Albert (uBA) and w3R family of graphs, we found that target layers  are $p_k=7,\ p_k=2$,  $p_k = 5$ and $ p_k=11$, respectively.

Using these targeted-single-layers, we did selective optimization using COBYLA optimizer and found out that for unweighted graph families, this parameters transfer initialization without optimization gives approximation ratios, $r_{n}\geq r_{GW}$. These value of $r_n$ is further improved  to $r_s$ by targeted-single layer optimization. This initialization was also able to produce $r_{n}\geq r_{GW}$ for w3R graph family but failed for wER family. Thus, full optimization is needed in wER family to achieve the approximation ratio, $r\geq r_{GW}\geq 0.878$.

In wBA family although $r_f$ were in the acceptable region ($\geq r_{GW}$) but the efficiencies of full optimization were similar to that of selective optimization. Thus it conclude that any one of the full and selective optimization can be used in wBA family but full optimization is slightly more efficient. 

We also see the effect of different optimizers in targeted selective optimization and found out that there were no significant difference in the accuracy but optimization times were different, concluding, COBYLA optimizer was most appropriate out of all other optimizers investigated. In the end, various initialization techniques were tested for full optimization and we found that parameters transfer initialization always achieved better accuracy in reduced optimization steps due to nearly optimal parameters transfer initialization.

We observe that for wER graph family, there is transition in $r$ value, using parameter transfer and targeted-single-layer optimization technique, that $r_{s}>r_{GW}$ for smaller nodes to $r_{s} <r_{GW}$ for higher nodes graph. One can further explore this transition more rigorously. Quantum loss landscape, defined in the main text by cost function, needs regularization for optimization. One could consider exploring the connection of Fourier transform loss function with different frequency components.  

\begin{acknowledgments}
U.M. acknowledges support from the Institute of Eminence (IoE),
University of Delhi, through the Faculty Research Program
(FRP) grants, sanctioned under Ref. No./IoE/2023-24/12/FRP and IoE/2024-25/12/FRP. SP thanks Dr. Vikash Chauhan for fruitful discussion, Dr. Rakesh Mishra and Dr. Subhajit Paul for allowing to access computational facility. SP thanks IBM for providing all the necessary libraries (Appendix \ref{app:appB}) for simulations.
\end{acknowledgments}

\appendix

\section{QAOA}
\label{app_qaoa}
\label{app:A}
QAOA is a hybrid quantum-classical algorithm to solve the COPs by applying $p$ layers of cost and mixer unitary alternatively, where cost unitary, $U_C(\gamma) = e^{-i\gamma H_C}$ encodes the problem and mixer unitary, $U_M(\beta) = e^{-i\beta H_M}$ explore the solution space to explore better solutions. The resulting quantum state is:
\begin{equation*}
	\ket{\psi(\gamma, \beta)} = \prod_{j=1}^{p} e^{-i\gamma_j H_C} e^{-i\beta_j H_M} \ket{+}^{\otimes n}.
\end{equation*}
where $\ket{+}^{\otimes n}$, denotes the uniform superposition of all computational basis states, $\gamma_j$ and $\beta_j$ are variational parameters need further classical optimization to minimize the expectation value of cost Hamiltonian:
\begin{equation*}
	(\gamma^*, \beta^*) = \arg\min_{{\gamma, \beta}} \bra{\psi(\gamma, \beta)}{H_C}\ket{\psi(\gamma, \beta)}.
\end{equation*}
It can achieve exact ground state energy when layers, $p\to \infty$. Practically, it is impossible to implement infinite number of layers, so an approximation can be taken by choosing intermediate number of layers.

\subsection*{MAXCUT}
\label{app_maxcut}
Given an undirected graph $G(V,E)$, where $V = \{1,2,...,n\}$ are number of vertices, $E$ denotes the edges and $w_{i,j}$ is the weight associated with each edge. The MaxCut problem is to find the subsets of vertices $S$ and $\bar S$ such that the weight of cut $(S, \bar S)$ defined by :
\begin{equation*}
	w(S, \bar S) = \sum_{i\in S, j\in \bar S} w_{i, j},
\end{equation*}
is maximum. This falls in the category of NP-hard problems. So solving it using classical computer is computationally challenging as the solution space grows exponentially by increasing input size, $N$. This problem equivalently can be converted into Ising glass problem as:
\begin{equation*}
	C(s) = \max\left(\frac{1}{2} \sum_{i,j \in E} w_{ij}(1 - s_is_j)\right), \quad \forall\ s_i \in \{-1, +1\}.
\end{equation*}

Converting equation into optimization problem obtained by changing $s_i \to z_i$ as:
\begin{equation*}
	H_C = \frac{1}{2} \sum_{i,j \in E} w_{ij}(\mathbb{I} - Z_iZ_j).
\end{equation*}
Where, $Z_i$ is Pauli-$z$ operators and $H_C$ represents the cost Hamiltonian. $\braket{H_C}$ is equivalent to $C(x)$ so ground state of $H_C$ corresponds to the optimal cut and performance is measured by the approximation ratio:

\begin{equation*}
	r(\gamma, \beta) = \frac{\bra{\psi(\gamma, \beta)}H_C\ket{\psi(\gamma, \beta)}}{C_{min}}.
\end{equation*}
where $C_{min}$ is the true ground state energy $H_C$ corresponds to the optimal cut. Where, $r=1$ when $p\to \infty$.

\section{Compute resources}
\label{app:appB}
All experiments were conducted using five systems each having specification: 13th Gen Intel® Core™ i9-13900K × 32, 5.8GHz CPU along with 128GB of RAM and 3TB of storage for the results operating on Ubuntu 24.04.3 LTS 64-bit.

Packages used in the simulations are following:\\
python = v3.12.12, numpy = v2.3.4, qiskit = v2.2.3, qiskit\_aer = v0.17.2, qiskit\_algorithms = v0.4.0, qiskit\_ibm\_runtime = v0.43.1, qiskit\_optimization = v0.7.0.

\bibliography{ref}




\end{document}